\documentclass{article}%
\usepackage{spconf,amsmath,epsfig}
\usepackage{amsmath,amsfonts,amssymb,graphicx,algorithm,color,cite,algorithmic}
\usepackage{amsfonts}
\usepackage{amssymb}
\usepackage{graphicx}
\usepackage{amsmath}%
\newtheorem{colloary}{Colloary}
\newtheorem{theorem}{Theorem}

\begin{document}

\title{\LARGE{Auction-based Resource Allocation for Multi-relay Asynchronous
Cooperative Networks}\vspace{-5mm}}
\name{Jianwei Huang, Zhu Han, Mung Chiang, and H. Vincent Poor\vspace{-10mm}}%
\vspace{-10mm}
\address{\thanks{J.~Huang is with the Department of Information Engineering, the Chinese University of Hong Kong, Shatin, NT, Hong Hong. Z.~Han is with the Electrical and Computer Engineering Department, Boise State University, Boise, ID, USA. M.~Chiang and H.V.~Poor are with the Department of Electrical Engineering, Princeton University, Princeton, NJ, USA. The work of J.~Huang is supported by Direct Grant of the Chinese Univ.~of Hong Kong under Grant 2050398. This work of H. V. Poor was supported by the U.S. National Science Foundation under Grants ANI-03-38807 and CNS-06-25637}
\vspace{-5mm}} \vspace{-10mm} \maketitle

\begin{abstract}
Resource allocation is considered for cooperative transmissions in
multiple-relay wireless networks. Two auction mechanisms, SNR
auctions and power auctions, are proposed to distributively
coordinate the allocation of power among multiple relays. In the
SNR auction, a user chooses the relay with the lowest weighted
price. In the power auction, a user may choose to use multiple
relays simultaneously, depending on the network topology and the
relays' prices. Sufficient conditions for the existence (in both
auctions) and uniqueness (in the SNR auction) of the Nash
equilibrium are given. The fairness of the SNR auction and
efficiency of the power auction are further discussed. It is also
proven that users can achieve the unique Nash equilibrium
distributively via best response updates in a completely
asynchronous manner.

\end{abstract}



\vspace{-3mm}

\section{Introduction}\vspace{-2mm}

Cooperative communication (e.g., \cite{sendonaris1}) takes
advantage of the broadcast nature of wireless channels, uses relay
nodes as virtual antennas, and thus realizes the benefits of
multiple-input-multiple-output (MIMO) communications in situations
where physical multiple antennas are difficult to install (e.g.,
on small sensor nodes). Although the physical layer performance of
cooperative communication has been extensively studied in the
context of small networks, there are still many open problems of
how to realize its full benefit in large-scale networks. For
example, to optimize cooperative communication in large networks,
we need to consider global channel information (including that for
source-destination, source-relay, and relay-destination channels),
heterogeneous resource constraints among users, and various upper
layer issues (e.g., routing and traffic demand). Recently some
centralized network control algorithms (e.g.,
\cite{NgYu07,Savazzi07}) have been proposed for cooperative
communications, but they require considerable overhead for
signaling and measurement and do not scale well with network size.
This motivates our study of distributed resource allocation
algorithms for cooperative communications in this paper.

In this paper, we design two distributed auction-based resource
allocation algorithms that achieve fairness and efficiency for
multiple-relay cooperative communication networks. Here fairness
means an allocation that equalizes the (weighted) marginal rate
increase among users who use the relay, and efficiency means an
allocation that maximizes the total rate increase realized by use
of the relays. Precise definitions of fairness and efficiency will
be given in Section
\ref{sec:sys_mod}. %
%
%
In both auctions, each user decides ``when to use relay'' based on a locally computable threshold
policy. The question of ``how to relay'' is answered by a simple weighted proportional allocation
among users who use the relay.

In our previous work \cite{Huang2007a}, we have proposed similar
auction mechanisms for a \emph{single-relay} cooperative
communication network, where users can achieve the desired auction
outcomes if they update their bids in a \emph{synchronous} manner.
This paper considers the more general case where there are
multiple relays in the network with different locations and
available resources. The existence, uniqueness, and properties of
the auction outcomes are very different from the single-relay
case. Moreover, we show that users can achieve the desirable
auction outcomes in a completely \emph{asynchronous} manner, which
is more realistic in practice and more difficult to prove. Due to
the space limitations, all the proofs are omitted in this
conference paper.

\vspace{-3mm}

\section{System Model and Network Objectives}
\label{sec:sys_mod}\vspace{-3mm}

\begin{figure}[tb]
\centering
\includegraphics[width=0.65\columnwidth, height=55truemm]{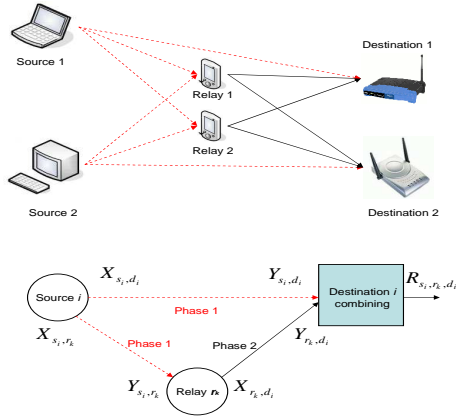}
\caption{System Model for Cooperation Transmission}%
\label{fig:system_model}\vspace{-8mm}%
\end{figure}

As a concrete example, we consider the  \emph{amplify-and-forward
(AF)} cooperative communication protocol in this paper. The system
diagram is shown in Fig. \ref{fig:system_model}, where there is a
set $\mathcal{K}=\left( 1,...,K\right) $ of relay nodes and a set
$\mathcal{I=}\left( 1,...,I\right) $ of source-destination pairs.
We also refer to pair $i$ as \textit{user }$i$, which includes
source node $s_{i}$ and destination node $d_{i}$.

For each user $i$, the cooperative transmission consists of two phases. In
\emph{Phase }$1$, source $s_{i}$ broadcasts its information with power
$P_{s_{i}}$. The received signals $Y_{s_{i},d_{i}}$ and $Y_{s_{i},r_{k}}$ at
destination $d_{i}$ and relay $r_{k}$ are given by $Y_{s_{i},d_{i}}%
=\sqrt{P_{s_{i}}G_{s_{i},d_{i}}}X_{s_{i}}+n_{d_{i}}$ and $Y_{s_{i},r_{k}%
}=\sqrt{P_{s_{i}}G_{s_{i},r_{k}}}X_{s_{i}}+n_{r_{k}}$, where $X_{s_{i}}$ is the
transmitted information symbol with unit energy at Phase $1$ at source $s_{i}%
$, $G_{s_{i},d_{i}}$ and $G_{s_{i},r_{k}}$ are the channel gains
from $s_{i}$ to destination $d_{i}$ and relay $r_{k},$
respectively, and $n_{d_{i}}$ and $n_{r_{k}}$ are additive white
Gaussian noises. Without loss of generality, we assume that the
noise level is the same for all links, and is denoted by
$\sigma^{2}$. We also assume that the transmission time of one
frame is less than the channel coherence time. The signal-to-noise
ratio (SNR) that is realized at
destination $d_{i}$ in Phase 1 is $\Gamma_{s_{i},d_{i}}=\frac{{P_{s_{i}%
}G_{s_{i},d_{i}}}}{{\sigma^{2}}}.$

In \emph{Phase }$2$, user $i$ can use a subset of (including all) relay nodes to help improve its
throughput. If relay $r_{k}$ is used by user $i$, $r_{k}$ will amplify $Y_{s_{i},r_{k}}$ and
forward it to destination $d_{i}$ with transmitted power $P_{r_{k},d_{i}}$. The received signal at
destination $d_{i}$ is
$Y_{r_{k},d_{i}}=\sqrt{P_{r_{k},d_{i}}G_{r_{k},d_{i}}}X_{r_{k},d_{i}}%
+n_{d_{i}}^{\prime},$where
$X_{r_{k},d_{i}}=Y_{s_{i},r_{k}}/|Y_{s_{i},r_{k}}|$ is the
unit-energy transmitted signal that relay $r_{k}$ receives from
source $s_{i}$ in Phase $1$, $G_{r_{k},d_{i}}$ is the channel gain
from relay $r_{k}$ to destination $d_{i}$, and
$n_{d_{i}}^{\prime}$ is the receiver noise in Phase $2$.
Equivalently, we can write $
Y_{r_{k},d_{i}}=\frac{\sqrt{P_{r_{k},d_{i}}G_{r_{k},d_{i}}}(\sqrt{P_{s_{i}%
}G_{s_{i},r_{k}}}X_{s_{i},d_{i}}+n_{r_{k}})}{\sqrt{P_{s_{i}}G_{s_{i},r_{k}%
}+\sigma^{2}}}+n_{d_{i}}^{\prime}. \label{eqn:dest signal}%
$
The additional SNR increase due to relay $r_{k}$ at $d_{i}$ is

\begin{equation}
\bigtriangleup\mathtt{SNR}_{ik}=\frac
{P_{r_{k},d_{i}}P_{s_{i}}G_{r_{k},d_{i}}G_{s_{i},r_{k}}}{\sigma^{2}%
(P_{r_{k},d_{i}}G_{r_{k},d_{i}}+P_{s_{i}}G_{s_{i},r_{k}}+\sigma^{2})}.
\label{eqn:relay SNR}%
\end{equation}
The total information rate user $i$ achieves at the output of
maximal ratio combining is
\begin{equation}
R_{s_{i},d_{i}}\left(  \boldsymbol{P}_{\boldsymbol{r},d_{i}}\right)
=\frac{W\log_{2}\left(  1+\Gamma_{s_{i},d_{i}}+\sum_{k}\bigtriangleup
\mathtt{SNR}_{ik}\right)  }{\sum_{k\in\mathcal{K}}\boldsymbol{1}_{\left\{
P_{r_{k},d_{i}}>0\right\}  }+1}. \label{eqn: MRC output}%
\end{equation}
Here $\boldsymbol{P}_{\boldsymbol{r},d_{i}}=\left(  P_{r_{k},d_{i}},\forall
k\in\mathcal{K}\right)  $ is the transmission power vector of all relays to
destination $d_{i}$, $W$ is the total bandwidth of the system, and
$\boldsymbol{1}_{\left\{  \cdot\right\}  }$ is the indicator function.
Equation (\ref{eqn: MRC output}) includes a special case where user $i$ does
not use any relay (i.e., $P_{r_{k},d_{i}}=0$ for all $k\in\mathcal{K}$), in
which case the rate is $W\log_{2}\left(  1+\Gamma_{s_{i},d_{i}}\right)  $. The
denominator in (\ref{eqn: MRC output}) models the fact that relay
transmissions occupy system resource (e.g., time slots, bandwidth, codes).
%
We write $R_{s_{i},d_{i}}\left(
\boldsymbol{P}_{\boldsymbol{r},d_{i}}\right) $ to emphasize that
$\boldsymbol{P}_{\boldsymbol{r},d_{i}}$ is the resource allocation
decision we need to make, and it is clear that $R_{s_{i},d_{i}}$
depends on other system parameters such as channel gains.

We assume that the source transmission power $P_{s_{i}}$ is fixed for each
user $i$. Each relay $r_{k}$ has a fixed total transmission power $P_{r_{k}}$,
and can choose the transmission power vector $\boldsymbol{P}_{r_{k}%
,\boldsymbol{d}}\triangleq( P_{r_{k},d_{1}},$
$...,P_{r_{k},d_{I}}) $ from the feasible set
\begin{equation}
\mathcal{P}_{r_{k}}\triangleq\left\{  \boldsymbol{P}_{r_{k},\boldsymbol{d}%
}\left\vert \sum_{i}P_{r_{k},d_{i}}\leq P_{r_{k}},P_{r_{k},d_{i}}\geq0,\forall
i\in\mathcal{I}\right.  \right\}  .
\end{equation}
Finally, define $\boldsymbol{P}_{\boldsymbol{r},\boldsymbol{d}}=\left(
\boldsymbol{P}_{r_{k},\boldsymbol{d}},\forall k\in\mathcal{K}\right)  $ to be the transmission
power of all relays to all users' destinations. The resource allocation decision we need to make is
the value of $\boldsymbol{P}_{\boldsymbol{r},\boldsymbol{d}}$.

From a network designer's point of view, it is important to
consider both \emph{efficiency} and \emph{fairness}. An efficient
power allocation
$\boldsymbol{P}_{\boldsymbol{r},\boldsymbol{d}}^{\text{efficiency}}$
maximizes the total rate increases of all users, i.e.,
\begin{equation}
\max_{\left\{  \boldsymbol{P}_{r_{k},\boldsymbol{d}}\in\mathcal{P}_{r_{k}%
},\forall k\in\mathcal{K}\right\}  }\sum_{i\in\mathcal{I}}\bigtriangleup
R_{i}\left(  \boldsymbol{P}_{\boldsymbol{r},d_{i}}\right)  ,
\label{Problem:efficiency}%
\end{equation}
where $\bigtriangleup R_{i}\left(  \boldsymbol{P}_{\boldsymbol{r},d_{i}%
}\right)  $ denotes the rate increase of user $i$ due to the use
of relays $ \bigtriangleup R_{i}\left(
\boldsymbol{P}_{\boldsymbol{r},d_{i}}\right)
=\max\left\{  R_{s_{i},d_{i}}\left(  \boldsymbol{P}_{\boldsymbol{r},d_{i}%
}\right)  -R_{s_{i},d_{i}}\left(  \boldsymbol{0}\right) ,0\right\}
$.
In many cases, an efficient allocation discriminates against users who are far
away from the relay. To avoid this, we also consider a fair power allocation
$\boldsymbol{P}_{\boldsymbol{r},\boldsymbol{d}}^{\text{fair}}$, where each
relay $r_{k}$ solves the following problem%
{\footnotesize
\begin{align}
\underset{\boldsymbol{P}_{r_{k},\boldsymbol{d}}\in\mathcal{P}_{r_{k}}}%
{\max}\;  \sum_{i}P_{r_{k},d_{i}}\label{Problem:Fairness},\ \mbox{s.t.}
\frac{\partial\bigtriangleup R_{i}\left( \bigtriangleup
\mathtt{SNR}_{ik}\right)  }{\partial\left(  \bigtriangleup\mathtt{SNR}%
_{ik}\right)
}=c_{k}q_{ik}\cdot\boldsymbol{1}_{\{P_{r_{k},d_{i}}>0\}},\forall
i\in\mathcal{I}.
\end{align}}
Here $q_{ik}$'s are the priority coefficients denoting the importance of each user to
each relay. When $q_{ik}=1$ for each $i$, all users who use relay $r_{k}$ have the same
marginal utility $c_{k}$, which leads to strict fairness among users.
In the special case where users are symmetric and only use the
same relay $r_{k}$, the fairness maximizing power allocation leads
to a Jain's fairness index \cite{Jain} equal to 1. However, the
definition of fairness here is more general than the Jain's
fairness index.
Notice that a fair allocation is Pareto optimal, i.e., no user's rate can be further
increased without decreasing the rate of another user.

Since $\bigtriangleup R_{i}\left(
\boldsymbol{P}_{\boldsymbol{r},d_{i}}\right)$ is non-smooth and
non-concave (due to the $\max$ operation), it is well known that
Problems (\ref{Problem:efficiency}) and (\ref{Problem:Fairness})
are $NP$ hard to solve even in a centralized fashion. Next, we
will propose two auction mechanisms that can solve these problems
under certain technical conditions in a distributed
fashion.\vspace{-2mm}

\vspace{-2mm}

\section{Auction Mechanisms}\vspace{-2mm}

\label{Sec: Share Auction}

An auction is a decentralized market mechanism for allocating
resources without knowing the private valuations of individual
users in a market. Auction theory has been recently used to study
various wireless resource allocation problems (e.g., time slot
allocation \cite{SunZheMod03} and power control
\cite{HuangMonet05} in cellular networks). Here we propose two
auction mechanisms for allocating resource in a multiple-relay
network. The rules of the two auctions are described below, with
the only difference being in payment determination.\vspace{-2mm}

\begin{itemize}
\item \emph{Initialization}: Each relay $r_{k}$ announces a
positive \textit{reserve bid} $\beta_{k}>0$ and a \textit{price}
$\pi_{k}>0$ to all users before the auction starts. \vspace{-2mm}

\item \emph{Bids}: Each user $i$ submits a nonnegative bid vector
$\boldsymbol{b}_{i}=\left(  b_{ik},\forall k\in\mathcal{K}\right)
$, one component to each relay.\vspace{-2mm}

\item \emph{Allocation}: Each relay $r_{k}$ allocates transmit
power as\vspace{-2mm}
\begin{equation}
P_{r_{k},d_{i}}=\frac{b_{ik}}{\sum_{j\in\mathcal{I}}b_{jk}+\beta_{k}}P_{r_{k}},\forall i\in\mathcal{I}. \label{eq:power-allocation}%
\end{equation}\vspace{-3mm}

\item \emph{Payments}: User $i$ pays $C_{i}=\sum
_{k}\pi_{k}q_{ik}\bigtriangleup\mathtt{SNR}_{ik}$ in an \emph{SNR} auction or
$C_{i}=\sum_{k}\pi_{k}P_{r_{k},d_{i}}$ in a power auction.\vspace{-2mm}
\end{itemize}
The two auction mechanisms that we propose are highly distributed, since each user only need to know the public system parameters (i.e., $W$, $\sigma^{2}$ and $P_{r_{k}}$ for all relay $k$), local information (i.e., $P_{s_{i}}$ and $G_{s_{i},d_{i}}$) and the channel gains with relays ($G_{s_{i},r_{k}}$ and $G_{r_{k},d_{i}}$ for
each relay $r_{k}$, which can be obtained through channel feedback). The relays do not need to know any network information.

A bidding profile is defined as the vector containing the users' bids, $\boldsymbol{b}=\left(
\boldsymbol{b}_{1},...,\boldsymbol{b}_{I}\right)  $.
The bidding profile of user $i$'s opponents is defined as $\boldsymbol{b}%
_{-i}=\left(  \boldsymbol{b}_{j},\forall j\neq i\right)  $, so that
$\boldsymbol{b}=\left(  \boldsymbol{b}_{i};\boldsymbol{b}_{-i}\right)  .$ User
$i$ chooses $\boldsymbol{b}_{i}$ to maximize its payoff
\begin{equation}
U_{i}\left(  \boldsymbol{b}_{i};\boldsymbol{b}_{-i},\boldsymbol{\pi}\right)
=\bigtriangleup R_{i}\left(  \boldsymbol{P}_{\boldsymbol{r},d_{i}}\left(
\boldsymbol{b}_{i};\boldsymbol{b}_{-i}\right)  \right)  -C_{i}\left(
\boldsymbol{b}_{i};\boldsymbol{b}_{-i},\boldsymbol{\pi}\right)  .
\end{equation}
Here $\boldsymbol{\pi}=\left(  \pi_{k},\forall k\in\mathcal{K}\right)  $ is the prices of
all relays. It can be shown that the values of the reserve bids $\beta_{k}$'s do not
affect the resource allocation, thus we can simply choose $\beta_{k}=1$ for all $k$.

The desirable outcome of an auction is called a \emph{Nash Equilibrium} (NE),
which is a bidding profile $\boldsymbol{b}^{\ast}$ such that no user wants to
deviate unilaterally, i.e.,
\begin{equation}
U_{i}\left(  \boldsymbol{b}_{i}^{\ast};\boldsymbol{b}_{-i}^{\ast
},\boldsymbol{\pi}\right)  \geq U_{i}\left(  \boldsymbol{b}_{i};\boldsymbol{b}%
_{-i}^{\ast},\boldsymbol{\pi}\right)  ,\forall i\in\mathcal{I},\forall
\boldsymbol{b}_{i}\geq0.
\end{equation}
Define user $i$'s \textit{best response} (for fixed $\boldsymbol{b}_{-i}$ and price
$\boldsymbol{\pi}$) as
\begin{equation}
\boldsymbol{\mathcal{B}}_{i}\left(  \boldsymbol{b}_{-i},\boldsymbol{\pi
}\right)  =\left\{  \boldsymbol{b}_{i}\left\vert \boldsymbol{b}_{i}=\arg
\max_{\boldsymbol{\tilde{b}}_{i}\geq\boldsymbol{0}}U_{i}\left(
\boldsymbol{\tilde{b}}_{i};\boldsymbol{b}_{-i},\boldsymbol{\pi}\right)
\right.  \right\}  ,\label{eq:best-response}%
\end{equation}
which can be written as $\boldsymbol{\mathcal{B}}_{i}\left(
\boldsymbol{b}_{-i},\boldsymbol{\pi}\right)  =\left(
\mathcal{B}_{i,k}\left(
\boldsymbol{b}_{-i},\boldsymbol{\pi}\right)  ,\forall
k\in\mathcal{K}\right) $. An NE is also a fixed point solution of
all users' best responses. Next we will consider the existence,
uniqueness and properties of the NE, and how to achieve it in
practice. Although in general NE is not the most desirable
operational point from an overall system point of view, we will
show later that the two auctions indeed achieve our desired
network objectives under suitable technical conditions.
\vspace{-2mm}

\subsection{SNR Auction}\vspace{-2mm}

We first consider the SNR auction where user $i$'s payment is
$C_{i}=\sum
_{k}\pi_{k}q_{ik}\bigtriangleup\mathtt{SNR}_{ik}.$\vspace{-2mm}

\begin{theorem}
\label{theorem: SNR_Auction_Multiple_Relay}In an SNR auction with
multiple relays, a user $i$ either does not use any relay, or uses
only one relay $r_{k(i)}$ with the smallest weighted price, i.e.,
$k(i)=\arg\min _{k\in\mathcal{K}}\pi_{k}q_{ik}$.\vspace{-2mm}
\end{theorem}

Theorem \ref{theorem: SNR_Auction_Multiple_Relay} implies that we
can divide a multiple-relay network into $K+1$ clusters of nodes:
each of the first $K$ clusters contains one relay node and the
users who use this relay, and the last cluster contains users that
do not use any relay. Then we can analyze each cluster
independently as a single-relay network as in \cite{Huang2007a}.
In particular, for a user $i$ belonging to cluster $k\left(
i\right) \leq K$, its best response function is
{\footnotesize\begin{equation}\label{eq:SNR-BR}
\mathcal{B}_{i,k}\left( b_{-i,k},\pi_{k}\right)  =\left\{
\begin{array}
[c]{cc}%
f_{i,k}^{s}\left(  \pi_{k}\right)  \left(  \sum_{j\neq i}b_{j,k}+\beta
_{k}\right),  & k=k\left(  i\right), \\
0, & \text{otherwise}.%
\end{array}
\right.
\end{equation}}
Note that user $i$'s best response is related only to the bids
from users who are in the same cluster. The linear coefficient
$f_{i,k}^{s}\left(  \pi
_{k}\right)  $ is derived as%
{\footnotesize
\begin{align}
&  f_{i,k}^{s}\left(  \pi_{k}\right)  =\\
&  \left\{
\begin{array}
[c]{cc}%
\infty, & \pi\leq\underline{\pi}_{i}^{s},\\
\text{{\small $\frac{\left(  P_{s_{i}}G_{s_{i},r_{k}}+\sigma^{2}\right)
\sigma^{2}}{\frac{P_{r_{k}  }G_{r_{k},d_{i}}P_{s_{i}}%
G_{s_{i},r_{k}}}{\frac{W}{2\pi_{k}q_{ik}\ln2}-1-{\Gamma_{s_{i},d_{i}}}%
}-\left(  P_{s_{i}}G_{s_{i},r_{k}}+P_{r_{k}}G_{r_{k},d_{i}}+\sigma^{2}\right)
\sigma^{2}},$}} & \pi\in\left(  \underline{\pi}_{i}^{s},\hat{\pi}_{i}%
^{s}\right), \\
0, & \pi\geq\hat{\pi}_{i}^{s},%
\end{array}\nonumber
\right.  %
\end{align}}
where
\begin{equation}
\underline{\pi}_{i}^{s}\triangleq\frac{W/\left(  2q_{ik}\ln2\right)
}{1+{\Gamma_{s_{i},d_{i}}+}\frac{P_{ r_{k} }G_{r_{k},d_{i}%
}P_{s_{i}}G_{s_{i},r_{k}}}{\left(  P_{s_{i}}G_{s_{i},r_{k}}+P_{r_{k}}G_{r_{k},d_{i}}+\sigma^{2}\right)  \sigma^{2}}},%
\end{equation}
and $\hat{\pi}_{i}^{s}$ is the \emph{smallest positive root} of the following
equation in $\pi$%
{\small
\begin{equation}
\pi q_{ik}\left(  1+{\Gamma_{s_{i},d_{i}}}\right)  -\frac{W}{2}\left(
\log_{2}\left(  \frac{2\pi q_{ik}\ln2}{W}\left(  1+{\Gamma_{s_{i},d_{i}}%
}\right)  ^{2}\right)  +\frac{1}{\ln2}\right)=0  . %
\end{equation}}
In the degenerate case where
$\hat{\pi}_{i}^{s}>\underline{\pi}_{i}^{s}$, we
have $f_{i,k}^{s}\left(  \pi_{k}\right)  =\infty$ for $\pi_{k}<\hat{\pi}%
_{i}^{s}$ and $f_{i,k}^{s}\left(  \pi_{k}\right)  =0$ for $\pi_{k}\geq\hat
{\pi}_{i}^{s}$. Notice that the linear coefficient is determined based on a simple
\emph{threshold} policy, i.e., comparing the price announced by the relay with the two
locally computable threshold prices.

Now let us assume that all users use the same relay $r_{k}$, then from
(\ref{eq:power-allocation}) and (\ref{eq:SNR-BR}) we know that the total demand for the
relay power is $\sum_{i\in\mathcal{I}}\frac{f_{i,k}^{s}\left(  \pi_{k}\right)  }%
{f_{i,k}^{s}\left(  \pi_{k}\right)  +1}P_{r_{k}}$, which can not exceed
$P_{r_{k}}$. It is also clear that $f_{i,k}^{s}\left(  \pi_{k}\right)  $ is a
non-increasing function of $\pi_{k}$. Then we can find a threshold price
$\pi_{k,th}^{s}$ such that $\sum_{i\in\mathcal{I}}\frac{f_{i,k}^{s}\left(
\pi_{k}\right)  }{f_{i,k}^{s}\left(  \pi_{k}\right)  +1}<1$ when $\pi_{k}%
>\pi_{k,th}^{s}$, and $\sum_{i\in\mathcal{I}}\frac{f_{i,k}^{s}\left(  \pi
_{k}\right)  }{f_{i,k}^{s}\left(  \pi_{k}\right)  +1}\geq1$ when $\pi_{k}%
\leq\pi_{k,th}^{s}$.

\begin{theorem}
\label{Theorem:SNR_existence}In an SNR auction with multiple relays, a unique NE exists if
$\pi_{k}>\pi_{k,th}^{s}$ for each $k$.\vspace{-2mm}
\end{theorem}

Finally let us consider the property of the NE. For a single-relay
network, we show in \cite{Huang2007a} that the SNR auction
achieves the fair resource allocation (i.e. it solves Problem
(\ref{Problem:Fairness})) if at least one user wants to use the
relay at the threshold price $\pi_{th}$. In the multiple-relay
case, however, some relays may never be able to achieve a Pareto
optimal allocation, which is a basic requirement for a fair
allocation. This is because if the relay announces a high price,
no users will use the relay. If the relay decreases the price,
there might be too many users switching to the same relay
simultaneously such that an NE\ does not exist. On the other hand,
we can show the following:\vspace{-2mm}
\begin{theorem}
If there exists a NE such that each relay's resource is full
utilized and each relay is used by at least one user, the
corresponding power allocation is fair (i.e., it solves Problem
(\ref{Problem:Fairness})).\vspace{-2mm}
\end{theorem}
\vspace{-3mm}

\subsection{Power Auction}\vspace{-2mm}

Here we consider the power auction, where user $i$'s payment is $C_{i}%
=\sum_{k}\pi_{k}P_{r_{k},d_{i}}$. There are two key differences
here compared with the SNR auction. First, a user may choose to
use multiple relays simultaneously here. User $i$'s best response
can be written in the following linear form: $
\mathcal{B}_{i,k}\left( b_{-i,k},\boldsymbol{\pi}\right)
=f_{i,k}^{p}\left( \boldsymbol{\pi}\right)$ $ \left(  \sum_{j\neq
i}b_{j,k}+\beta_{k}\right) ,\forall k\in\mathcal{K}. $ To
calculate $f_{i,k}^{p}\left( \boldsymbol{\pi}\right)  $, user $i$
needs to consider a total of $\sum_{l=0}^{K}\binom{K}{l}$ cases of
choosing relays. For example, when there are two relays in the
network, a user needs to consider four cases: not using any relay,
using relay $1$ only, using relay $2$ only, and using both relays.
For the given relay choice in case $n$, it calculates the linear
coefficients $f_{i,k}^{p,n}\left( \boldsymbol{\pi }\right) $ for
all $k$ in closed-form (this involves threshold policy similar to
the SNR auction) and the corresponding rate increase
$\bigtriangleup R_{i}^{n}$ . Then it find the case that yields the
largest payoff, $n^{\ast}=\arg\max_{n}\bigtriangleup R_{i}^{n}$,
and sets $f_{i,k}^{p}\left( \boldsymbol{\pi }\right)
=f_{i,k}^{p,n^{\ast}}\left( \boldsymbol{\pi}\right)  $ $\forall
k$. Second, the linear coefficient $f_{i,k}^{p}\left(
\boldsymbol{\pi}\right)  $ depends on the prices announced by all
relays. For example, either a large $\pi_{k}$ or a small
$\pi_{k^{\prime}}$ ($k^{\prime}\neq k$) can make
$f_{i,k}^{p}\left( \boldsymbol{\pi}\right) =0$, i.e., user $i$
will choose not to use relay $r_{k}$.

Similar to in the SNR auction, we can also calculate a threshold
price $\pi_{k,th}$ for relay $r_{k}$. In this case, we assume that
all relays announce infinitely high prices except $r_{k}$, and
then calculate $\pi _{k,th}^{p}$ such that
$\sum_{i\in\mathcal{I}}\frac{f_{i,k}^{s}\left(  \pi
_{k}\right)  }{f_{i,k}^{s}\left(  \pi_{k}\right)  +1}<1$ when $\pi_{k}%
>\pi_{k,th}^{p}$, and $\sum_{i\in\mathcal{I}}\frac{f_{i,k}^{s}\left(  \pi
_{k}\right)  }{f_{i,k}^{s}\left(  \pi_{k}\right)  +1}\geq1$ when $\pi_{k}%
\leq\pi_{k,th}^{p}$.\vspace{-2mm}
\begin{colloary}
In a power auction with multiple relays, there exists an NE if
$\pi_{k}>\pi_{k,th}^{p}$ for each $k$.\vspace{-2mm}
\end{colloary}
On the other hand, necessary condition for existence of NE as well
as conditions for uniqueness are not straightforward to specify,
and are left for future research. We can characterize the property
of the NE as follows:\vspace{-2mm}

\begin{theorem}
If there exists a NE such that each relay's resource is full
utilized and all users use all relays, the corresponding power
allocation is efficient (i.e., it solves Problem
(\ref{Problem:efficiency})).\vspace{-2mm}
\end{theorem}


\vspace{-3mm}

\subsection{Asynchronous Best Response Updates}\vspace{-3mm}

\label{sect:distributed_algorithm}

The last question we want to answer is how the NE can be reached
in a distributed fashion. Since user $i$ does not know the best
response functions of other users, it is impossible for it to
calculate the NE in one shot. In the context of a single-relay
network \cite{Huang2007a}, we have shown that distributed best
response updates can globally converge to the unique NE (if it
exists) in a \emph{synchronous} manner, i.e., all users update
their bids in each time slot simultaneously accordingly to $
b_{i}\left( t\right) =f_{i}^{s}\left(  \pi\right)  \left(
\sum_{l\neq i}b_{l}\left(  t-1\right) +\beta\right). $ In
practice, however, it would be difficult or even undesirable to
coordinate all users to update their bids at the same time, and
the following can be used:

\begin{algorithm}
\caption{Asynchronous Best Response Bid Updates}
\begin{algorithmic} [1]
\label{algo1}%

\STATE$t=0$.

\STATE Each user $i$ randomly chooses a $\boldsymbol{b}_{i}\left(
0\right)  \in\left[  \underline{\boldsymbol{b}}_{i},\boldsymbol{\bar{b}}%
_{i}\right]  .$

\STATE$t=t+1.$\label{line:increase time}

\STATE\textbf{for} each user $i\in\mathcal{I}$

\STATE$\ \ \ \ $\textbf{if } $t\in\mathcal{T}_{i}$ \textbf{then}

\STATE$\ \ \ \ \ \ \ \ b_{i,k}\left(  t\right)  =\left[ f_{i}^{s}\left( \boldsymbol{\pi}\right)
\left(  \sum_{l\neq i}b_{l}\left(  t-1\right) +\beta\right) \right]
_{\underline{b}_{i,k}}^{\bar{b}_{i,k}},\forall k$. \label{line:modified_best_response}

\STATE$\ \ \ \ $\textbf{end if}

\STATE\textbf{end for}

\STATE Go to Step \ref{line:increase time}.%

\end{algorithmic}
\end{algorithm}%

We show that \emph{asynchronous} best response updates converges
in the \emph{multiple-relay} case. The complete asynchronous best
response update algorithm is given in Algorithm \ref{algo1}
($\left[  x\right] _{a}^{b}=\max$ $\left\{ \min\left\{ x,b\right\}
,a\right\}  $.), where each user $i$ updates its bid only if the
current time slot belongs to a set $\mathcal{T}_{i}$, which is an
unbounded set of time slots and could be different from user to
user. We make a very mild assumption that the asynchronism of the
updates is bounded, i.e., there exists a finite but sufficiently
large positive constant $B$, and for all
$t_{1}\in\mathcal{T}_{i},\ $there exists a
$t_{2}\in\mathcal{T}_{i}\ $such that $t_{2}-t_{1}\leq B$. Each
user updates its bid at least once during any time interval of
length $B$ slots. The exact value of $B$ is not important (as long
as it is bounded) for the convergence proof  and needs not to be known by the users.%
\vspace{-2mm}

\begin{theorem}
\label{Theorem:convergence} If there exists a unique nonzero NE in
the SNR auction, there always exists a lowerbound bid vector
$\underline{\boldsymbol{b} }\boldsymbol{=}$ $\left(
\boldsymbol{\bar{b}}_{i},\forall i\in\mathcal{I} \right)  $ and an
upperbound bid vector $\boldsymbol{\bar{b}=}\left(
\underline{\boldsymbol{b}}_{i},\forall i\in\mathcal{I}\right)  $,
under which Algorithm \ref{algo1} globally converges to the unique
NE.\vspace{-2mm}
\end{theorem}

In practice, we can choose  $\underline{\boldsymbol{b}}$ to be a sufficiently small
positive vector (to approximate zero bids from users) and $\boldsymbol{\bar{b}}$ to be a
sufficiently large finite vector.

\vspace{-3mm}
\section{Simulation Results}\vspace{-3mm}

\label{sec:simulation}

\begin{figure}[t]
\begin{minipage}[b]{0.48\columnwidth}
\centering
\centerline{\epsfig{figure=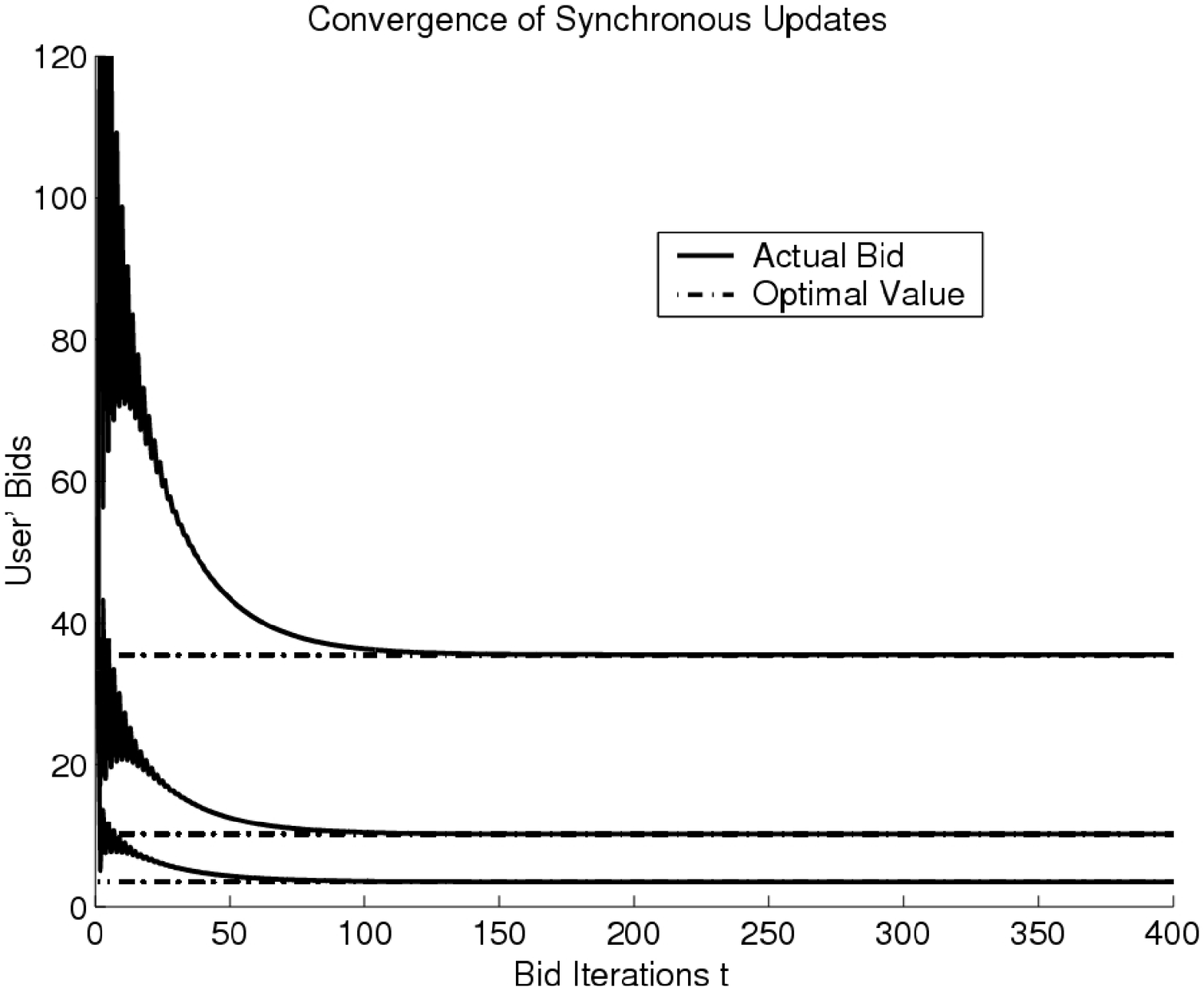,width=\columnwidth}}
\centerline{\small{(a) Synchronous Updates}}
\end{minipage}%
\hfill%
\begin{minipage}[b]{0.5\columnwidth}
\centering
\centerline{\epsfig{figure=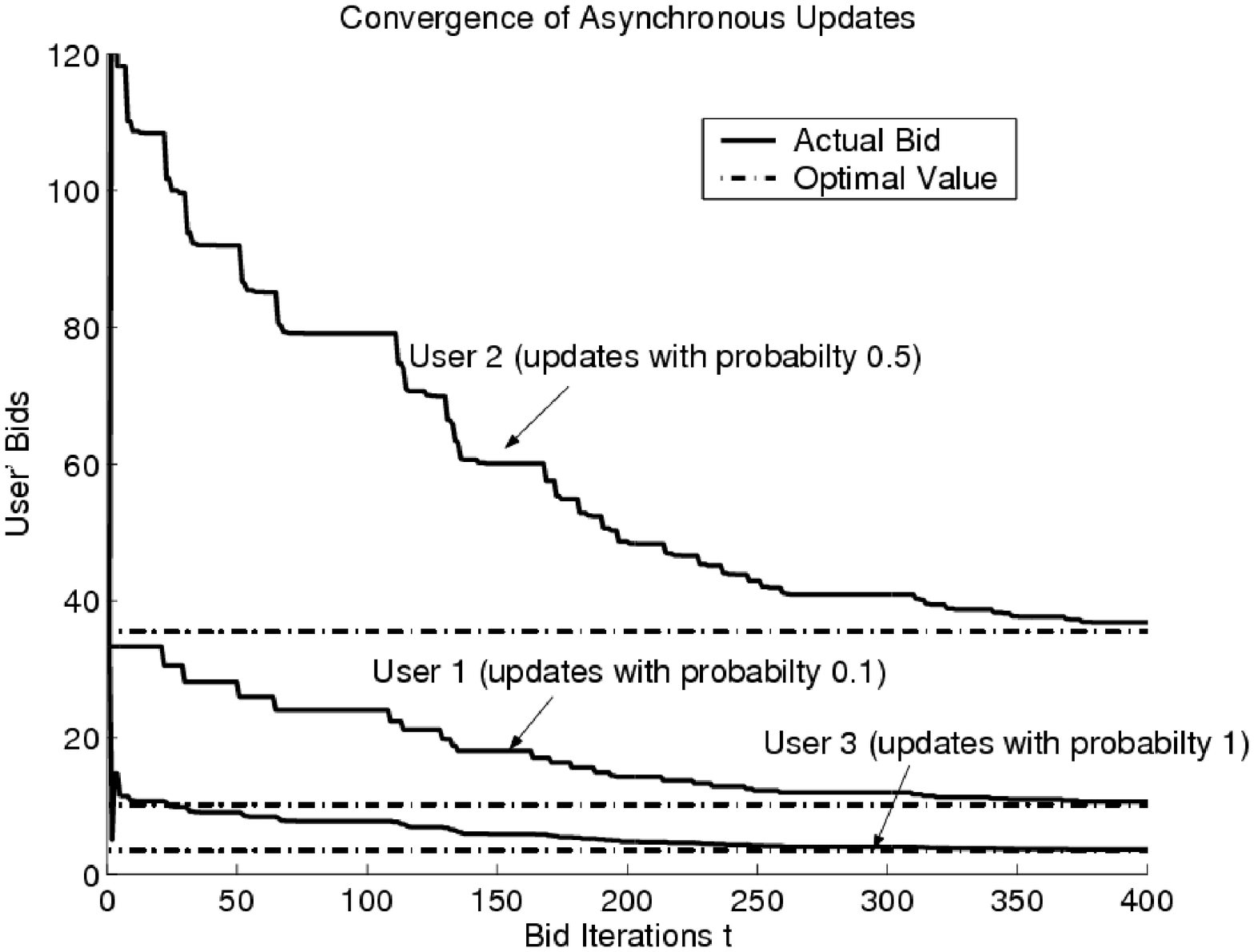,width=\columnwidth}}
\centerline{\small{(b) Asynchrony Updates}}
\end{minipage}
\caption{Bids update in an SNR auction (the same one relay).}
\label{fig:update}\vspace{-5mm}
\end{figure}

For illustration purpose, we show the convergence of Algorithm
\ref{algo1} in a multiple-relay SNR auction. We consider a network
with three users and two relays. The three transmitters are
located at (100m,-25m), (-100m,25m) and (100m,5m), and the three
receivers are located at (-100m,25m), (100m,25m) and (-100m,5m).
The two relays are located at (0m,-2m) and (0m,0m). All the
priority coefficients $q_{ik}=1$. Since the first relay announces
a price lower than the second relay, all users choose to use the
first relay. In Fig.~\ref{fig:update}.a, we show the convergence
of the users' bids to the first relay under synchronous updates,
where
each user updates its bid in each time slot. The solid lines show
the evolution of the bids and the dotted lines show the optimal
values of the bids after convergence. In Fig.~\ref{fig:update}.b,
we show the convergence under the same setup with asynchronous
convergence. Three users randomly and independently choose to
update their own bids in each time slot with probability $0.1$,
$0.5$ and $1$, respectively. We can see that the algorithm
converges to the same optimal values as the synchronous update
case but in longer time (as expected).

\vspace{-3mm}

\section{Conclusions}\vspace{-3mm}

\label{sec:conclusions}


In this paper, a cooperative communication network with multiple
relays has been considered, and two auction mechanisms, the SNR
auction and the power auction, have been proposed to
distributively coordinate the relay power allocation among users.
Unlike the single-relay case studied in \cite{Huang2007a}, here
the users' choices of relays depend on the prices announced by all
relays. In the SNR auction, a user will choose the relay with the
lowest weighted price. In the power auction, a user might use
multiple relays simultaneously, depending on the network topology
and the relative relationship among the relays' prices. A
sufficient condition is shown for the existence of the Nash
equilibrium in both auctions, and conditions are derived for
uniqueness in the SNR auction. The fairness of the SNR auction and
the efficiency of the power auction are also discussed. Finally,
if an NE exists, users can achieve it in a distributed fashion via
best response updates in an asynchronous manner.

\end{document}